\documentclass[conference]{IEEEtran}
\usepackage{cite}

\ifCLASSINFOpdf
  \usepackage[pdftex]{graphicx}
  \graphicspath{{../fig/}{./fig/}}
  \DeclareGraphicsExtensions{.png,.jpeg,.pdf}
\else
  \usepackage[dvips]{graphicx}
	\DeclareGraphicsExtensions{.eps}
\fi
\usepackage[caption=false,font=footnotesize]{subfig}

\usepackage{url}

\usepackage{multirow}
\usepackage{booktabs}
\usepackage{colortbl}
\usepackage{paralist}

\newcommand{\formatpar}[1] {
  \begin{minipage}{3.0cm}
	\textbf{#1} 
	\end{minipage}
}

\newcommand{\cellmod}{
	\cellcolor[gray]{.9}
}

\usepackage[shadow]{todonotes}

\usepackage{listings} 
\lstnewenvironment{notes}{
\lstset{%
  basicstyle=\ttfamily\color{gray},  
  breaklines=true,
  columns=fullflexible,
  escapeinside = ||,
  breakindent=0.5cm,
  inputencoding={utf8},
  moredelim=**[is][\itshape]{*}{*},
  literate=%
    {Ö}{{\"O}}1
    {Ä}{{\"A}}1
    {Ü}{{\"U}}1
    {ß}{{\ss}}1
    {ü}{{\"u}}1
    {ä}{{\"a}}1
    {ö}{{\"o}}1
    {~}{{\textasciitilde}}1
  }
}{}

\begin{document}
%
\title{Does Quality of Requirements Specifications matter?\\Combined Results of Two Empirical Studies}

\author{\IEEEauthorblockN{Jakob Mund, Henning Femmer, Daniel M\'{e}ndez Fern\'{a}ndez, Jonas Eckhardt}
\IEEEauthorblockA{Technische Universit\"{a}t M\"{u}nchen\\
Garching b. M\"{u}nchen, Germany\\
\texttt{\{mund,femmer,mendezfe,eckharjo\}@in.tum.de}}
}



%


\maketitle

\begin{abstract}

[\textit{Background}] 
Requirements Engineering is crucial for project success, and to this end, many measures for quality assurance of the software requirements specification (SRS) have been proposed. 
[\textit{Goal}]
However, we still need an empirical understanding on the extent to which SRS are created and used in practice, as well as the degree to which the quality 
of an SRS matters to subsequent development activities.
[\textit{Method}]
We studied the relevance of SRS by relying on survey research and explored the impact of quality defects in SRS by relying on a controlled experiment.
[\textit{Results}]
Our results suggest that the relevance of SRS quality depends both on particular project characteristics and what is considered as a quality defect; for instance, the domain of safety critical systems seems to motivate for an intense usage of SRS as a means for communication whereas defects hampering the pragmatic quality do not seem to be as relevant as initially thought.
[\textit{Conclusion}]
Efficient and effective quality assurance measures must be specific for carefully characterized contexts and carefully select defect classes.

\end{abstract}

\IEEEpeerreviewmaketitle

\section{Introduction} 

Stake\-holder-appropriate requirements constitute critical determinants of project quality. Incorrect or missing requirements are supposed to lead to various problems in later phases such as effort and time overrun or an increased effort in acceptance testing~\cite{MW14}. In fact, a large extent of documented project failures are meant to be caused by insufficient requirements engineering (RE)~\cite{MW14,kamata2007does,kujala2005role,hofmann2001requirements}. 

It became conventional wisdom that the quality of the created RE artifacts, most prominently the software requirements specification (SRS), are a measurement for the overall process quality as weaknesses in the SRS might cause problems in subsequent development phases. In literature, we can consequently find various proposals on how to structure an SRS including content models~(e.g., \cite{fernandez2014artefact}), and broader documentation guidelines as well as best-practices~(e.g.,\cite{wiegers1999writing,ieee830-1998}), and finally quality models~(e.g., \cite{berry2006new,ott2012defects,krogstie1995towards,pohl1994three,femmer15abreqm}) on what characteristics an SRS should feature, all together supposed to improve the quality in RE processes. The role and relevance of RE artifacts thereby became a frequent field of investigation (see e.g.~\cite{liskin2015artifacts}). 

However, yet not completely answered is the question how much the process quality is eventually determined by the quality of the artifacts, which includes also the question how much project participants eventually rely on the created artifacts. Consequently, we are confronted with the following interesting and, at the same time, challenging questions
\begin{compactitem}
\item under which (project) circumstances an SRS matters, 
\item what quality dimensions of an SRS matter, and finally 
\item how we can properly assure the quality of an SRS.
\end{compactitem}
Since the explicit documentation of requirements and their quality assurance are labor-intensive tasks, practitioners are often confronted with a trade-off between effort and adherence to schedules on the one hand and the achievement of the necessary quantity and quality of requirement documentation on the other hand. 

\textbf{Problem Statement:} So far, we lack an empirical understanding on the extent to which SRS are created and used in practice, as well as the degree to which the quality of an SRS matters to subsequent development activities that rely on the artifacts. Such an understanding would also allow for a critical reflection on the effectiveness of SRS-based quality assurance.

\textbf{Research Objectives:} We aim at contributing to a better understanding on the impact the quality of an SRS eventually has and formulate two research objectives to understand:
\begin{compactenum}
\item[RO~1:] To which extent and under which conditions are SRS created and used?
\item[RO~2:] Does the quality of an SRS matter to subsequent development activities?
\end{compactenum}

\textbf{Contribution:} In the paper at hands, we make two contributions:
\begin{compactenum}
\item We conduct a survey to explore to which extent and under which conditions SRS are created and used in practical environments. This contribution shall address RO~1.
\item We design and execute a controlled experiment to analyse the degree to which the quality of an SRS matters to subsequent development activities. This contribution shall address RO~2.\end{compactenum}
The results of our survey are presented in Sect.~\ref{sec:survey}, the results of our experiment are presented in Sect.~\ref{sec:experiment}. Based on our findings, we  critically reflect on the challenges introduced above regarding SRS-based quality assurance in RE in Sect.~\ref{sec:discussion}.

\section{Related Work}
\label{sec:related_work}
Several studies address (the use of) \emph{documentation} of requirements engineering in practice. 
Liskin investigated, among other, in a qualitative study~\cite{liskin2015artifacts}, the suitability of specific RE artifacts for
activities related with requirements specifications, by means of interviews.
%
%
Lethbridge, Singer and Forward~\cite{lethbridge2003software} reported on three empirical studies on documentation in software engineering in practice.
Results identified both applications of documentation in general during software engineering and issues with documentation, in particular, out-dated information.

Furthermore, there exists empirical evidence providing insights into how requirements are \emph{communicated}. 
Abelein and Paech~\cite{abelein2014state} conducted a series of semi-structured interviews concerning the state of the practice of user-developer communication in large-scale IT projects. The results of their study indicate that the direct user-developer communication is limited and that no common method for this communication in the design and implementation method exist. We extend the context of their work by adding multiple stakeholders (i.e., users of the SRS) but focus on the SRS as the single means for communication.
Bjarnason et al.~\cite{bjarnason2011requirements} conduct an explanatory case study in order to deepen their understanding of the cause and effects of communication gaps in a large-scale industrial setup. Their results show that that communication gaps cause failure to meet the customers' expectations, quality issues, and wasted effort. In contrast to our work, their study is of explanatory nature, and furthermore has a larger scope (communication gaps in general).

In summary, we refine the existing body of knowledge by specifically challenging the relevance of the documentation (quality), taking into account project-specific circumstances and certain classes of qualities, 
in order to provide a more subtle picture on the impact of SRS quality in practice.

\section{Relevance of SRS: An Expert Survey}
\label{sec:survey}

To answer the question overall about the whether and how much the quality of SRS matter, we must first clarify if it is actually used. To this end, we conducted a survey with a broad spectrum of practitioners from one industrial partner to explore the extent to which SRS are created and used. 

\subsection{Research Questions}

In this study, we explore two facets of an SRS, namely its degree of completeness and detail in the documentation of requirements in a persistent artifact and its use as a means for communicating requirements within RE and beyond RE. To this end, for formulate two research questions described next.

\subsubsection*{RQ\,1-1: To which extent are requirements documented?}

This research questions examines the degree to which requirements are documented in SRS or comparable artifacts (e.g., product backlogs). We distinguish between two dimensions when documenting requirements. First, we want to know how comprehensive requirements are documented in terms of quantity, i.e. what is the proportion of documented requirements compared to all requirements identified during requirements engineering. Second, requirements can be specified with varying level of detail. Therefore, we want to know how detailed the requirements are documented.


\subsubsection*{RQ\,1-2: To what degree are SRS used to communicate requirements?}

SRS, or comparable artifacts, also serve the purpose to communicate requirements from stakeholders to various roles in the systems engineering process, e.g., architects, implementers or testers. RQ~2 investigates the degree to which internalization of knowledge about requirements is based on the SRS. 
To this end, we want to know whether and how often a SRS is used as a means for communication considering both the communication within RE and the communication of requirements to subsequent development activities. In case it is used, we further want to know as how important it is seen in comparison with other means of communicating requirements in practice.

\subsubsection*{Is SRS usage related to specific project circumstances?}

In a secondary study, Kalus and Kuhrmann~\cite{kalus2013criteria} identified criteria which 
lead to (i)~an expansion resp.\ reduction of documentation within projects, and (ii)~an orientation towards formalized resp.\ open communication patterns. Since we expect the SRS to not be used equally under all circumstances, we want to know in addition to both research questions if there are specific criteria which influence whether requirements are documented~(\emph{RQ~1-1}) and a SRS is used for communicating requirements~(\emph{RQ~1-2}). We therefore use the secondary study of Kalus and Kuhrmann~\cite{kalus2013criteria} as a basis for our hypotheses against which we test our data samples.

\subsection{Survey Design}
The survey was conducted at a large, multi-national company headquartered in Germany. 
Although operating in different domains, typical products are medium to large systems or engineering solutions in which software plays a significant or even crucial role. 

\subsubsection*{Participants} We targeted participants directly or indirectly involved in requirements engineering for software-intensive systems, either in the sense of being involved when eliciting and specifying requirements, or in the sense of relying with their particular activities on requirements, e.g. architects or implementers. 

\subsubsection*{Survey Instrument} The questionnaire consists of two main parts. Part \textsc{I} includes questions on the frequency of project characteristics to occur independent of individual projects, and part \textsc{II} refers to the most recently completed project the participant was involved in. 
For that particular project, the participants are asked to characterize (a)~the project itself, (b)~the extent to which requirements are documented in a SRS and/or (c)~the use of the SRS as a means to communicate requirements. Questions of part \textsc{II}(b) and \textsc{II}(c) are only shown if the participant specified she was involved in the elicitation and specification of requirements respectively required knowledge of requirements for her tasks, in the particular project. In the questionnaire, we relied mostly on closed questions, with open questions to capture rationales or unforeseen options, e.g., means of communications. The presence of project characteristics (\textsc{II}-a) was captured by Lickert-scales ranging from 1 (e.g. \emph{``I strongly disagree''}) to a maximum of 6 (e.g. \emph{``I strongly agree''}) to avoid checking the middle, while the extent of documentation (\textsc{II-b}) as perceived by participants was captured on 5-point ordinal scales (cf.\ Fig.~\ref{fig:survey_results_rq1}). Details on the instrument are available online\footnote{\url{www4.in.tum.de/~mund/srs-quality-om.zip}}.

\subsubsection*{Data Collection} We implemented the questionnaire as an online survey 
using the \textit{Enterprise Feedback Suite 10.5} tool. Due to organizational restrictions, we only
conducted the survey anonymously. We made the survey available to participants working in systems engineering project via an announcement on selected working-group mailing lists of the company. In addition, we selected participants based on former, company-internal projects conducted in collaboration with our partner. However, the list of participants was undisclosed to us. 

\subsubsection*{Data Analysis}
For RQ~1-1, we considered only those participants who stated to be involved in requirements elicitation or specification. For both the completeness and the level of detail, we extracted the number of projects for each level of the ordinal scales. For RQ~1-2, we compared the number of projects in which an SRS (paper-based and tool-based) was used to those of meetings/workshops, personal talks, and groupware solutions. Furthermore, for those respondents who stated to use an SRS, we evaluated the participants ranking of the communication means according to specified relevance for informing about requirements. 

We investigated the relation between the presence of project criteria and the usage of the SRS by means of rank-based correlation coefficients and hypothesis test based on Kendall's $\tau$. 
The null hypothesis is that the presence of a project criterion (part \textsc{II}(a), 4 point Lickert-scale) and usage of the SRS (see Fig.~1 and 2 for scales) are not correlated. We rated $\tau \geq 0.3$ as moderate and $\tau \geq 0.5$ as strong correlation and used a significance level of $\alpha = .05$. Because we tested multiple hypothesis, we also calculated an adjusted p-value $p_{fdr}$ based on Benjamini and Hochberg~\cite{benjamini1995controlling} to mitigate the threat that correlations were only found by chance. In addition, we extracted reasons from open questions for complete, incomplete, shallow and detailed specifications, which we then either assigned to existing criteria or generalized as candidates for new criteria.

\subsubsection*{Validity Procedures}
\label{sec:survey_validity_procedures}

The instrument for our survey was reviewed by two additional researchers and two industrial partners 
who also checked no terms with different or ambiguous meanings 
within the company were used. To avoid both bias towards single projects and multiple answers, 
we verified all projects gathered in the survey are unique by manually comparing the 
specified project names (mandatory). Qualitative data resulting from open questions was reviewed independently  to avoid misinterpretation. For identified correlations (RQ~1\,\&\,2), 
results were visualized using bubble plots and checked for plausibility.

\subsection{Results and Interpretation}
In the following, we summarize our results structured according to the research questions. For each, we conclude with a brief interpretation of the results.

\subsubsection*{Study Population}

The survey was accessed 85 times, of which 46 participants (54\%) completed the survey\footnote{61 participants (72\%) partly completed the survey, mostly until the mandatory specification of the project name}. Four participants did specify to neither being involved during requirements specification nor that knowledge on requirements was required for their tasks. The remaining 42 participants had an average experience of $\geq$10 years and completed $6-10$ projects on average.
\begin{table}[!t]

\renewcommand{\arraystretch}{1.3}

\caption{Participants by Project Role and RE Involvement (During Specification or in the sense of Required Knowledge of Requirements)}
\label{tbl:survey_pop_participants}
\centering
\begin{tabular}{@{}p{2.5cm}ccc@{}} \toprule
  \textbf{Role} & \textbf{Specification} \ & \textbf{Required Knowledge (excl.)} & \textbf{Total} \\ 
	\midrule
	
	Product Manager				& 9 & 4 (-) & 9 \\
	Project Lead					& 5 & 4 (-) & 5 \\
	Req. Engineer					& 9 & 5 (1) & 10 \\
	Architect							& 5 & 8 (3) & 8 \\
	Implementer						& - & 4 (4) & 4 \\
	Tester								& 1 & 1 (-) & 1 \\
	Quality Manager				& 1 & 2 (1) & 2 \\
	Other									& 1 & 3 (2) & 3 \\
	\midrule
	All & 31 & 31 (11) & 42 \\
	\bottomrule
\end{tabular}
\end{table}
Projects specified by the participants were balanced between products and custom solutions (23 to 19)
and between new and continuous development (22 to 20). The majority of projects (71\%) had release cycles between six month and two years, but short ($\leq$6 months) and long ($\geq$5 years) also occurred.

%
%
\subsubsection*{RQ\,1: Documenting requirements in SRS}
\label{sec:survey_results_rq1}

\begin{figure*}[!t]

\subfloat[Degree of Completeness]{\includegraphics[width=0.33\textwidth]{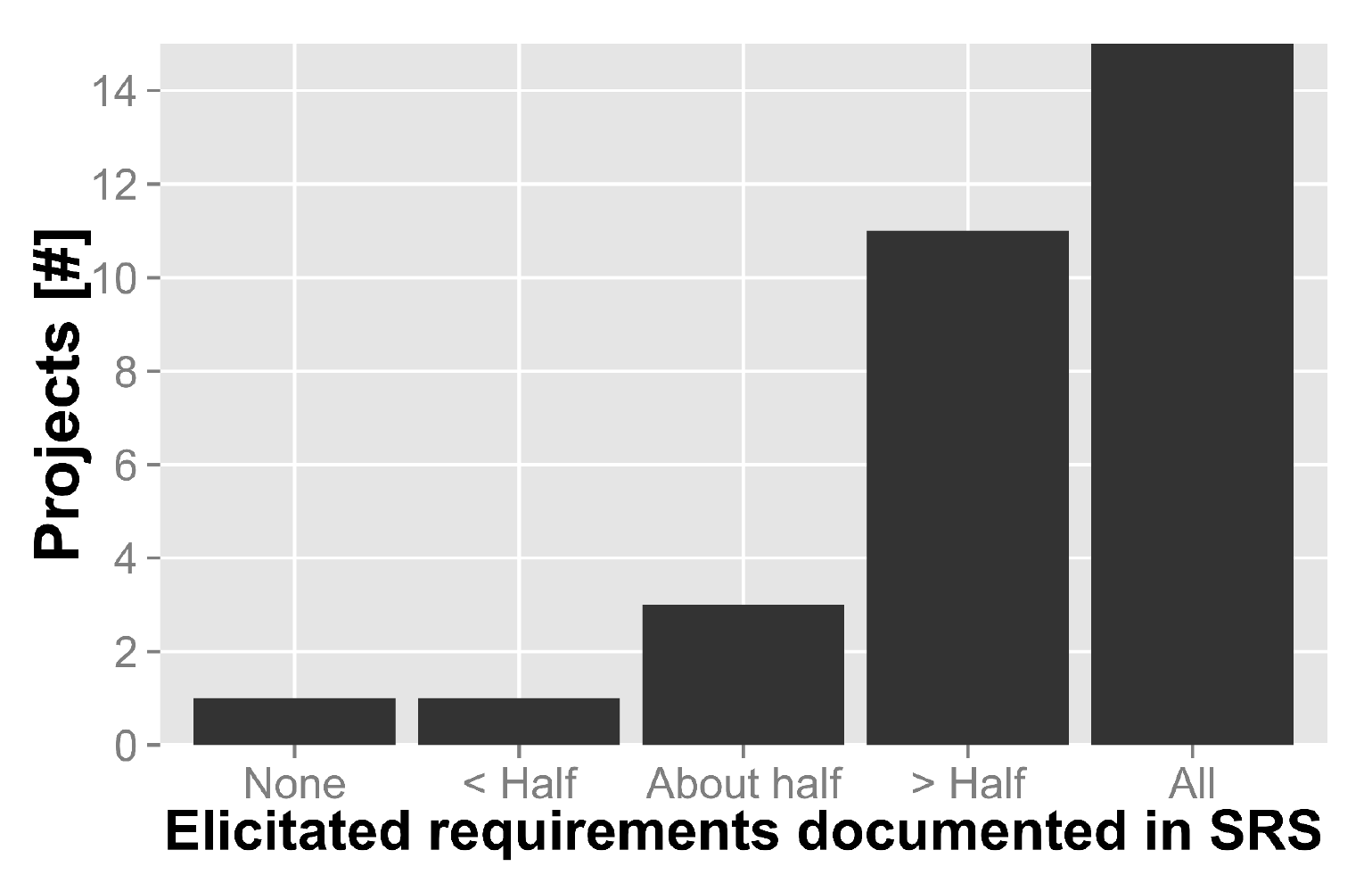} \label{fig:survey_i111}}
\subfloat[Level of Detail]{\includegraphics[width=0.33\textwidth]{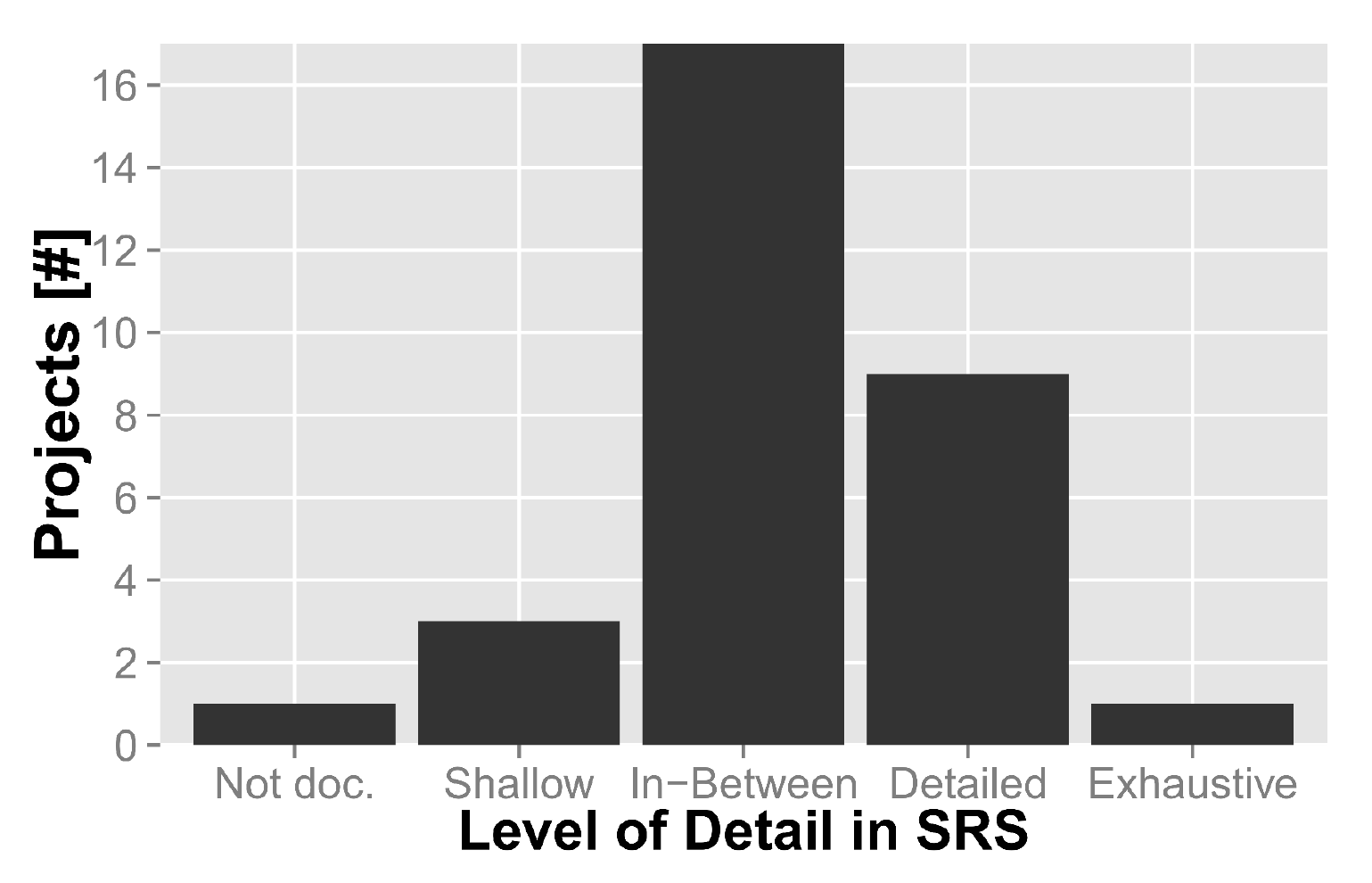} \label{fig:survey_i121}}
\subfloat[Correlation of (a) and (b)]{\includegraphics[width=0.33\textwidth]{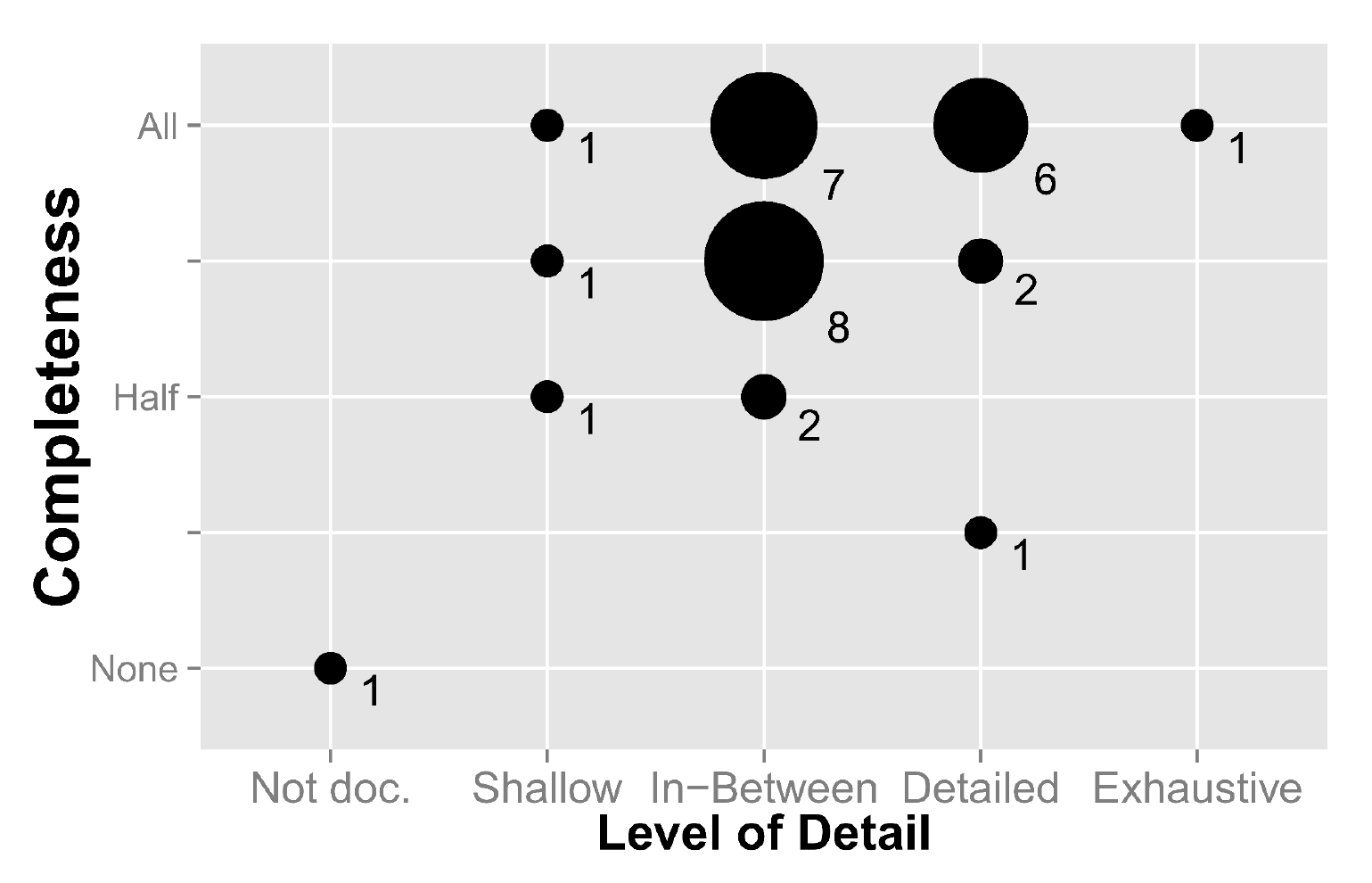} \label{fig:survey_corr}}

\caption{Survey results on requirements documentation (RQ1): portion of documented requirements~(a), associated level of detail~(b) and correlation~(c)}
\label{fig:survey_results_rq1}
\end{figure*}

In general, our results indicate to a rather comprehensive usage of SRS to document requirements. Considering the scope, the respondents stated for 15 out of 31 projects that all identified requirements were actually documented in a SRS, and in only four cases (13\%) half or less of identified requirements were documented (Fig.~\ref{fig:survey_results_rq1}\subref*{fig:survey_i111}). The level of detail of the documented requirements was balanced between shallow and detailed specifications (17 cases, 55\%). SRS were considered shallow (3 cases, 10\%) and exhaustively detailed (1 case, 3\%) only rarely (Fig.~\ref{fig:survey_results_rq1}\subref*{fig:survey_i121}). We found a moderate rank-correlation (Kendall-$\tau=0.33$, $p=0.04$) between the degree of completeness and the level of detail in SRS, depicted as a bubble chart in Fig.~\ref{fig:survey_results_rq1}\subref*{fig:survey_corr}.

The relationship between project criteria and the documentation, i.e., the degree of completeness and the level of detail, is described by the rank-correlation coefficients and associated $p$-values in the left part of Tab.~\ref{tbl:survey_coef}. The data shows positive correlations between the completeness and three factors, namely that 
\begin{compactenum}
\item safety/security concerns are relevant for the project goal
\item measurements are required (both project-specific and overall)
\item the team and stakeholders work in a good and collaborative way (project-specific only)
\end{compactenum}

All but the last also positively correlate with the level of detail in the SRS, but the project-specific correlation regarding measurements was not statistically significant. In contrast, a high complexity of the system under consideration (project-specific and overall) and volatile requirements (overall only) correlate negatively with the completeness of the SRS. The circumstance that the stakeholders and the team worked together in previous cooperation negative correlations with the level of details in an SRS.

Qualitative feedback supports the visible correlations. Several participants stated that the degree of completeness in the SRS was required by the development process, with multiple participants stating that it is a consequence of the domain (\emph{regulatory requirements (healthcare sector) enforce documentation}) and/or required for ensuring traceability (which is perceived as \emph{mandatory for healthcare products} and \emph{required for distributed teams}.
 
In agile processes, SRS documents were perceived as complete due to \emph{only documented requirements entering the backlog}. Moreover, participants stated that complete requirement specifications were obtained due to the \emph{involvement of all (relevant) stakeholders} and \emph{iterations between development and product management}. In contrast, bare \emph{[existence of] too many requirements (thousands)} or the application of \emph{traditional RE approaches on complex projects} resulted in incomplete requirements specification according. Long release cycles were also perceived as a reason either directly, since \emph{topics [\dots] which are relevant late were documented very coarsely} and \emph{long project durations [imply] many changes}, or more indirectly, because \emph{permanent changing goals, constraints, stakeholders and project teams}. Also, external documentation (e.g., \emph{availability of legacy systems}), limited time (\emph{restrictive time-boxing} and \emph{not enough time [\dots] to elicit all requirements}), and \emph{stakeholder lacked knowledge of requirements in detail} were mentioned.

For the level of detail, participants mentioned that \emph{good coordination of requirements in the team allowed for less detail in documentation} and a \emph{rough direction [is] enough [because] experts clarify details during implementation}. Most predominantly, participants stated solution-orientation as a reason for a shallow level of detail including statements like (\emph{results more important than documentation}, \emph{urgency for technical results limits time for specifications}) and \emph{not enough time and resources for detailed analysis}. \emph{Development process constraints} were mentioned as an exemplary reason for detailed SRS.

\paragraph*{Interpretation} {In general, requirements seem to be not exhaustively documented in an SRS in every project. Our results suggest that requirements are, however, documented in nearly every project, and with substantial quantity (completeness) and a high level of detail. Since the results indicate to a 
higher exhaustiveness regarding the completeness than the level of specification detail (prevalence of projects above an imaginary diagonal in Fig.\ref{fig:survey_corr}), we conclude so far that the former is more important than the latter. If we take specific project circumstances into account, we can observe that correlations obtained for individual projects are generally weaker than for the overall set of projects the respondents worked in. This may indicate that documentation is influenced to a large degree not by individual project circumstances, but by the chosen domain-specific development process. 

Based on the revealed correlations and a priori hypothesis~\cite{kalus2013criteria}, we conclude that safety/security concerns and demand for measurement increase the need for documentation, and propose a novel hypothesis that a good cooperation between stakeholders (principal) and agent allows more exhaustive documentation, and vice versa. For negative correlations, we argue that volatile requirements hamper the degree of completeness of SRS. However, we are indecisive 
if high complexity decreases the need for documentation, because of the inherent inefficiency associated with the difficulties of documenting such requirements (\emph{uselessness of traditional RE approaches}), or whether it simply impedes documentation without actually diminishing its need.
}

%
%
\subsubsection*{RQ\,2: SRS as Communication Means}

\begin{figure*}[!t]
\caption{Usage of SRS: means for communicating requirements ranked according to the participants' perceived importance for information}
\label{fig:survey_results_rq2}
\centering
\subfloat[Usage \label{fig:survey_rq2_media}]{\includegraphics[width=0.30\textwidth]{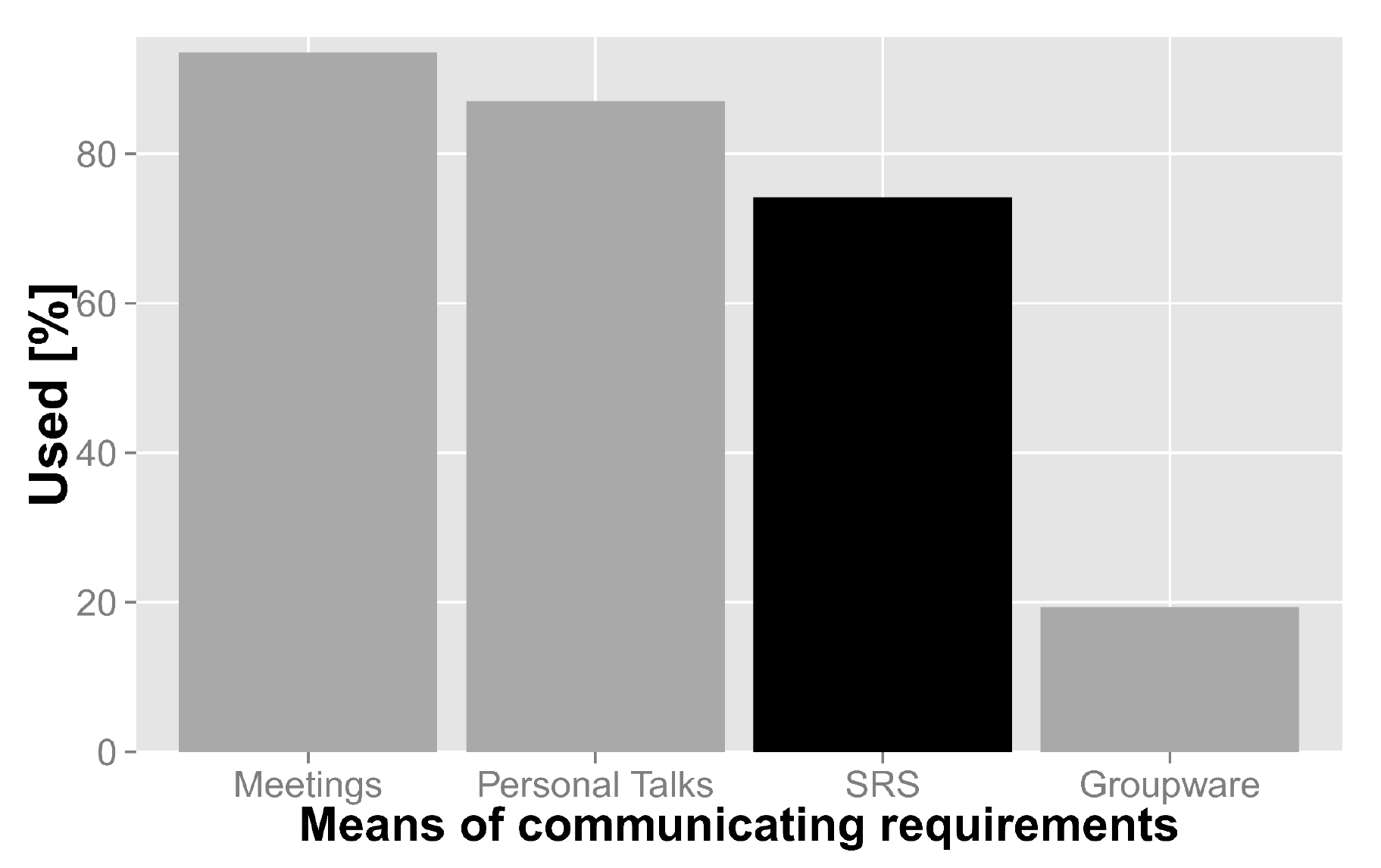} }
\subfloat[Ranking \label{fig:survey_rq2_ranking}]{\includegraphics[width=0.30\textwidth]{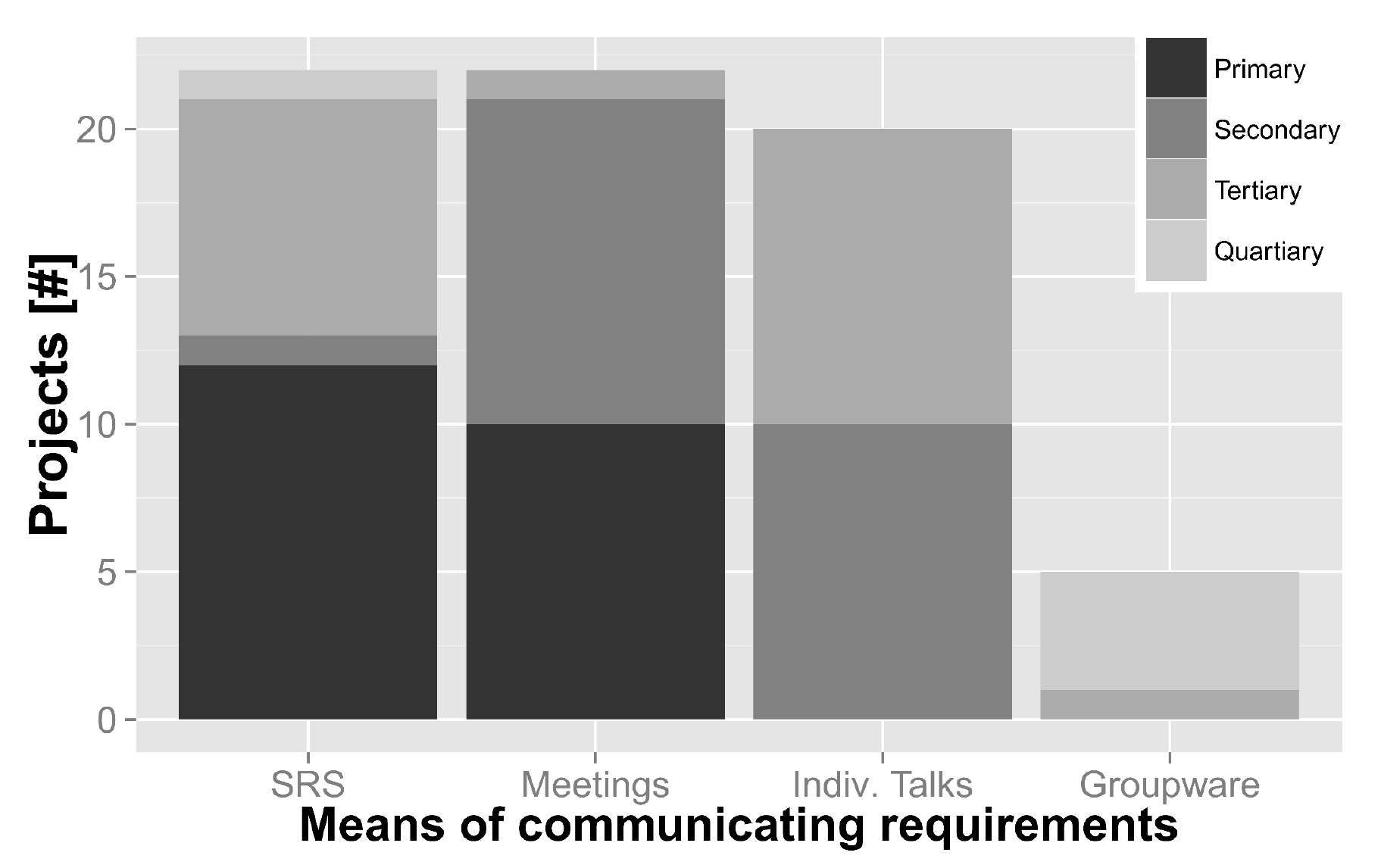}}
\end{figure*}

The SRS, independent of its materialization (paper-based and tool-based), was used as a means to communicate requirements in 23 out of 31 projects (74\%), ranked third behind meetings/workshops (29 projects, 94\%) and personal talks (27 projects, 87\%, cf.~Fig.~\ref{fig:survey_results_rq2}\subref*{fig:survey_rq2_media}). Considering participants not involved during specification exclusively, the SRS is used slightly more often (82\%, $+8\%$), while meetings/workshops (91\%, $-3\%$) and personal talks (82\%, $-5\%$) are used marginally less often. In any case, groupware solutions are used rarely (20\% and 27\%, respectively).
To further investigate the importance of the SRS as a communication means, we asked the participants to rank the specified means by how informative they were perceived for their individual project tasks. Our results reveal that in case an SRS is used, it is the primary source for communicating requirements (55\%), but only slightly more often than meetings/workshops (45\%). In fact, we observed a pronounced polarisation between SRS-based and artifact-agnostic communication. If meetings are the primary source, individual personal talks were specified predominantly as the secondary source of information, with the SRS being used in only one case as the secondary means. Out of five projects using groupware solutions, it was considered the least in all but one case, where it was ranked second to last, attesting groupware solutions an inferior relevance for communication. Considering only participants not involved during the requirements specification, SRS-based communication of requirements was considered as primary source significantly more often (75\%, $+20\%$), indicating its superiority for communicating requirements for development phases subsequent to requirements engineering.

Compared to the documentation of requirements, we observed a more alleviated effect of project characteristics on the communication of requirements. First and foremost, we could not reject the null hypothesis for the principal usage of an SRS, and identify only two correlations for the ranking of communication means: a strong correlation ($\tau$=0.52, $p$=0.01) with the relevancy of safety/security for the project goal, and a moderate correlation with the length of release cycles ($\tau$=0.33, $p$=0.08). However, our results show several weak correlations (around $\tau$=0.2) which are in tune with our hypothesis. For instance, we found weak correlations for team parameters regarding size and turnover, as well as some of the factors identified as significant for the documentation (cf.~Sec.~\ref{sec:survey_results_rq1}), such as the demand for measurements ($\tau$=0.27 for ranking) or high complexity ($\tau$=~--0.18 for usage and $\tau$=--~0.12 for ranking). In addition, we investigated whether the participants' background knowledge regarding the domain or product impacts the communication. However, we could not reject the null hypothesis, and hence background knowledge may not have an impact on the communication means used\footnote{Note that due to scoping, we did not investigate whether there is \emph{less} communication, independent of the actual \emph{means} to transfer knowledge}.

\begin{table*}[!htb]
\renewcommand{\arraystretch}{1.3}

\caption{Impact of Project Parameters on the Relevance of the SRS (moderate corr.\ for $\tau\geq.3$, sig.\ level $\alpha=.05$)}
\label{tbl:survey_coef}
\centering
\begin{tabular}{@{}p{3.0cm}rrrrlrrrlrrrlrrr@{}} \toprule
  \multicolumn{2}{c}{\multirow{3}{*}{\textbf{Parameter}}} 	& \multicolumn{7}{c}{\textbf{Impact on Documentation in SRS}} 																				& \phantom{a} &  \multicolumn{7}{c}{\textbf{Impact on Communication using SRS}} \\
	\multicolumn{2}{c}{}			 											&	\multicolumn{3}{c}{Completeness} 		& \phantom{a}	& \multicolumn{3}{c}{Level of Detail} 		& & \multicolumn{3}{c}{SRS used?} 	& \phantom{a}				& \multicolumn{3}{c}{SRS Ranking} \\ 
	\cmidrule{3-5} \cmidrule{7-9} \cmidrule{11-13} \cmidrule{15-17}
	\multicolumn{2}{c}{}														&	$\tau$~~ & $p$~~ & $p_{fdr}$~ && $\tau$~~ & $p$~~ & $p_{fdr}$~ && $\tau$~~ & $p$~~ & $p_{fdr}$~ && $\tau$~~ & $p$~~ & $p_{fdr}$~ \\ \midrule


\multirow{2}{*}{\formatpar{Team Size}} & Proj.\ &-0.08 & 0.63 & 0.85 && 0.06 & 0.72 & 0.95 && 0.07 & 0.65 & 0.8 && 0.2 & 0.28 & 0.63 \\
& Gen.\ & -0.08 & 0.6 & 0.82 && 0.17 & 0.28 & 0.52 && 0.04 & 0.81 & 1 && 0.05 & 0.79 & 0.89 \\ \addlinespace 

\multirow{2}{*}{\formatpar{Team Distribution}} & Proj.\ &0.06 & 0.7 & 0.85 && 0.02 & 0.89 & 0.95 && 0.09 & 0.59 & 0.8 && 0.02 & 0.92 & 0.92 \\
& Gen.\ & -0.04 & 0.81 & 0.84 && 0.25 & 0.12 & 0.32 && -0.14 & 0.41 & 1 && 0.09 & 0.62 & 0.85 \\ \addlinespace 

\multirow{2}{*}{\formatpar{Team Turnover}} & Proj.\ &-0.02 & 0.88 & 0.93 && 0.06 & 0.71 & 0.95 && -0.01 & 0.94 & 0.94 && 0.22 & 0.26 & 0.63 \\
& Gen.\ & -0.09 & 0.56 & 0.82 && 0.01 & 0.94 & 0.95 && -0.04 & 0.83 & 1 && 0.19 & 0.32 & 0.75 \\ \addlinespace 

\multirow{2}{*}{\formatpar{Management unavailable}} & Proj.\ &0.01 & 0.93 & 0.93 && 0.06 & 0.69 & 0.95 && 0.1 & 0.54 & 0.8 && 0.13 & 0.52 & 0.73 \\
& Gen.\ & 0.03 & 0.84 & 0.84 && 0.28 & 0.08 & 0.29 && 0.08 & 0.61 & 1 && 0.26 & 0.19 & 0.75 \\ \addlinespace 

\multirow{2}{*}{\formatpar{Financial controlling req.}} & Proj.\ &0.09 & 0.59 & 0.85 && 0.12 & 0.44 & 0.95 && 0.15 & 0.37 & 0.77 && 0.12 & 0.53 & 0.73 \\
& Gen.\ & -0.05 & 0.73 & 0.84 && 0.09 & 0.55 & 0.75 && 0.3 & 0.07 & 0.71 && -0.18 & 0.34 & 0.75 \\ \addlinespace 

\multirow{2}{*}{\formatpar{Measurement req.}} & Proj.\ &\cellmod 0.34 & \cellmod 0.03 & 0.18 && 0.27 & 0.1 & 0.53 && 0.14 & 0.4 & 0.77 && 0.27 & 0.17 & 0.62 \\
& Gen.\ & \cellmod 0.37 & \cellmod 0.02 & 0.08 && \cellmod 0.37 & \cellmod 0.02 & 0.16 && 0.25 & 0.13 & 0.71 && -0.21 & 0.29 & 0.75 \\ \addlinespace 

\multirow{2}{*}{\formatpar{Many stakeholders}} & Proj.\ &-0.1 & 0.55 & 0.85 && 0.06 & 0.71 & 0.95 && -0.05 & 0.77 & 0.85 && 0.04 & 0.84 & 0.92 \\
& Gen.\ & -0.11 & 0.51 & 0.82 && 0.16 & 0.33 & 0.52 && 0.14 & 0.42 & 1 && -0.15 & 0.46 & 0.8 \\ \addlinespace 

\multirow{2}{*}{\formatpar{Stakeholder unavailable}} & Proj.\ &-0.12 & 0.46 & 0.85 && 0.03 & 0.87 & 0.95 && 0.21 & 0.22 & 0.77 && 0.08 & 0.69 & 0.84 \\
& Gen.\ & -0.17 & 0.29 & 0.64 && 0.17 & 0.28 & 0.52 && 0 & 1 & 1 && -0.03 & 0.89 & 0.89 \\ \addlinespace 

\multirow{2}{*}{\formatpar{High complexity}} & Proj.\ &\cellmod -0.31 & 0.06 & 0.21 && 0.01 & 0.95 & 0.95 && -0.18 & 0.31 & 0.77 && -0.12 & 0.53 & 0.73 \\
& Gen.\ & \cellmod -0.36 & \cellmod 0.02 & 0.08 && 0.02 & 0.92 & 0.95 && 0.03 & 0.86 & 1 && 0.03 & 0.89 & 0.89 \\ \addlinespace 

\multirow{2}{*}{\formatpar{Safety/Security relevant}} & Proj.\ &\cellmod 0.38 & \cellmod 0.02 & 0.18 && \cellmod 0.36 & \cellmod 0.02 & 0.25 && 0.14 & 0.42 & 0.77 && \cellmod 0.52 & \cellmod 0.01 & 0.08 \\
& Gen.\ & \cellmod 0.32 & \cellmod 0.04 & 0.12 && \cellmod 0.34 & \cellmod 0.03 & 0.16 && 0.01 & 0.96 & 1 && \cellmod 0.38 & \cellmod 0.05 & 0.51 \\ \addlinespace 

\multirow{1}{*}{\formatpar{Release Cycle Length}} & Proj.\ &-0.19 & 0.24 & 0.67 && -0.01 & 0.95 & 0.95 && 0.26 & 0.11 & 0.77 && \cellmod 0.33 & 0.08 & 0.46
\\ 

 \midrule 

\multirow{2}{*}{\formatpar{Previous cooperation}} & Proj.\ &-0.03 & 0.83 & 0.83 && \cellmod -0.33 & \cellmod 0.04 & 0.16 && -0.11 & 0.52 & 0.81 && -0.07 & 0.71 & 0.81 \\
& Gen.\ & -0.01 & 0.95 & 0.95 && -0.19 & 0.22 & 0.45 && 0.03 & 0.87 & 0.96 && -0.17 & 0.38 & 0.81 \\ \addlinespace 

\multirow{2}{*}{\formatpar{Good cooperation}} & Proj.\ &\cellmod 0.44 & \cellmod 0.01 & \cellmod 0.04 && 0.19 & 0.25 & 0.34 && 0.07 & 0.71 & 0.81 && -0.05 & 0.81 & 0.81 \\
& Gen.\ & 0.26 & 0.11 & 0.23 && -0.15 & 0.36 & 0.48 && -0.07 & 0.69 & 0.96 && -0.29 & 0.15 & 0.81 \\ \addlinespace 

\multirow{2}{*}{\formatpar{Small Budget}} & Proj.\ &0.12 & 0.45 & 0.6 && -0.19 & 0.23 & 0.34 && 0.04 & 0.81 & 0.81 && -0.14 & 0.48 & 0.81 \\
& Gen.\ & 0.01 & 0.95 & 0.95 && -0.27 & 0.08 & 0.32 && 0.18 & 0.28 & 0.96 && 0.02 & 0.92 & 0.81 \\ \addlinespace 

\multirow{2}{*}{\formatpar{Volatile Requirements}} & Proj.\ &-0.21 & 0.19 & 0.38 && -0.04 & 0.81 & 0.81 && -0.08 & 0.62 & 0.81 && -0.1 & 0.59 & 0.81 \\
& Gen.\ & \cellmod -0.47 & \cellmod 0 & \cellmod 0.01 && -0.01 & 0.95 & 0.95 && -0.01 & 0.96 & 0.96 && -0.13 & 0.51 & 0.81 \\

 \midrule 

\multirow{1}{*}{\formatpar{Prev. Domain Knowledge}} & Proj.\ &- & - & - && - & - & - && 0.28 & 0.09 & 0.18 && 0 & 1 & 1 \\ \addlinespace
\multirow{1}{*}{\formatpar{Prev. Product Knowledge}} & Proj.\ &- & - & - && - & - & - && -0.03 & 0.85 & 0.85 && -0.08 & 0.68 & 1 \\ 


\bottomrule											
\end{tabular}
\end{table*}

\paragraph*{Interpretation}{Overall, we conclude so far that SRS are a well-established means to communicate requirements. However, as we cannot guarantee that our participants reflect the composition of project teams in practice, we focus on a distinction between two important communication relationships, namely a communication within RE activities, e.g. elicitation and negotiation, and a communication of RE results to subsequent development activities, e.g. testing. We rely our interpretation on the observation that participants involved during the specification of requirements exhibit significantly different communication preferences than participants who only required knowledge of the requirements for their individual project assignments. Consequently, we argue that for projects with a high degree of division of labor in terms of project activities, communication within RE and adjacent activities (e.g., high-level architecture, cf.~Tab.~\ref{tbl:survey_pop_participants}) non-artifact-based communication means prevail, while for the communication of SRS to subsequent development activities, an SRS seems predominately used for that communication. 

In contrast to the documentation of requirements, the use of SRS as a communication means may be less determined by the development processes but be specific to the project or even the individual. The later could also explain why it was only possible to reveal weak correlations, e.g., team size ($\tau$=0.20) and distribution ($\tau$=0.22). For moderate or strong correlations, we propose the following causal interpretations and possible explanations: the length of release cycles impacts the use of SRS for communication (e.g., because of the persistent nature of artifacts), and that safety/security concerns demands documented traceability. Also, we interpret that the degree of documentation effects the role of the SRS for communication, supported by the revealed moderate correlation between the degree of documentation and its use for communication ($\tau$=0.36 for completeness, and $\tau$=0.46 for level of detail), but limited to participants involved in specification \emph{and} requiring knowledge of requirements for their project tasks.\footnote{Since we relied on the specified completeness and level of detail as perceived by participants during specification.}} 

In summary, we therefore draw the conclusion that the SRS is created in detail and with a high degree of completeness under specific project circumstances (such as the application domain) to communicate requirements.

\subsection{Limitations}

Considering the internal validity, we had to cope with limited control regarding sampling and delivery because of the particular industrial setting. Hence, we were unable to establish a random sampling or accomplish higher response rates. Therefore, statistical results have to be considered with a salt of grain and participation bias is possible. While we used the survey results to gain first insights into when and how the SRS is used, the number of participants was too low to apply statistical methods reliably when discriminating between aspects, e.g. for the subgroup of participants not involved in the specification of requirements. 

Also, we cannot conclude with statistical significance that the observed correlations occurred only by chance, since most of the corrected confidence intervals $p_{fdr}$ were above the threshold. Despite our validity procedures, we suspect some terms still to be subject to misinterpretation, partly because of the heterogeneity of requirements engineering in practice.

\section{Impact of Quality Defects in SRS: A Controlled Experiment}
\label{sec:experiment}

Our study in Sect.~\ref{sec:survey} concludes that in different project situations SRS are used to foster communication with subsequent activities. Here, we investigate the complementary question to what extend the quality of the RE artifacts, if used, actually impacts subsequent engineering activities. Since a complete analysis of all subsequent activities is infeasible in a controlled setting, our study focuses on an activity that strongly depends on the contents of SRS: system testing. We further focus on two deficiencies of SRS, an exemplary one for semantic quality and another one for pragmatic quality (relying on the terminology of Lindland et al.~\cite{lindland1994understanding}).

\subsubsection*{RQ\,2-1: Do incorrect SRS statements impact system testing?}
According to \cite{lindland1994understanding}, semantic quality defects can be characterized as incomplete and/or incorrect information in the SRS with respect to the stakeholder's actual demands on the system.
In RQ\,2-1, we investigate whether incorrect information in the SRS inevitable leads to flawed system test cases or makes the inference of system tests less efficient.

\subsubsection*{RQ\,2-2: Do negative SRS statements impact system testing?}
In contrast, RQ\,2-2 focuses on pragmatic quality, i.e.\ the unambiguous comprehensibility of the SRS by the target audience, e.g., test engineers. Such defects (cf.\ ISO29148~\cite{ISO2011} for a list) describe valid information but can nonetheless lead to flawed test cases if the SRS is misunderstood or not understood at all. In a previous investigation~\cite{femmer14smells}, practitioners were unsure about the validity of \emph{negative statements}, one quality factor in this list, yet without empirical foundation. Therefore, we investigate the impact of this particular defect on system testing in terms of omissions or incorrect test cases.

\subsubsection*{RQ\,2-3: Does domain knowledge compensate for quality defects in SRS?}
The main confounding factor that prevents generalization of results from controlled settings is context knowledge. 
Hence, we want to know if and to what degree a-priori knowledge about the application domain, also called the problem space~\cite{jackson2001problem}, effectively compensates the quality defects of \emph{RQ\,2-1} and \emph{RQ\,2-2}.

\begin{table*}[!t]
\renewcommand{\arraystretch}{1.3}

\caption{Experiment treatment: Presence of defects (y/n) injected into use cases of SRS}
\label{tbl:exp_defects}
\centering
\begin{tabular}{@{}lllp{6.7cm}p{6.3cm}@{}} \toprule
  \textbf{Use Case} & \textbf{Defect} & \textbf{Type} \ & \textbf{Correct statement} & \textbf{Flawed statement} \\ 
	\midrule
	
	\multirow{2}{*}{UC 1} & D1.1			 & Incorrect & \emph{In case the data is complete and valid, the data will be imported.} & \emph{In case the data is complete and valid, \textbf{an error message occurs}.} \\
										 & D1.2		  & Incorrect & \emph{Data records must then be further approved
										by a case handler and explicitly activated.} & \emph{Data records must then be checked by a \textbf{plausibility algorithm 
										and are activated afterwards}.} \\
	\midrule
	\multirow{3}{*}{UC 2} & D2.1  & Negation & \emph{The user must enter at least one character.} & \emph{The user \textbf{is not allowed to enter zero characters.}} \\
	& D2.2  & Negation & \emph{The system must treat lowercase and uppercase letters the same.} & \emph{The system \textbf{must not 
	distinguish} between lowercase and uppercase letters.} \\
	& D2.3  & Negation & \emph{The user may only select one company at a time.} & \emph{The user \textbf{cannot select more than one company} at a time.} \\
	
	\midrule
	
	\multirow{2}{*}{UC 3} & D3.1 & Negation & \emph{The user may nominate up to three substitutes.} & \emph{\textbf{No user must nominate more than} three substitutes.} \\
	
	& D3.2  & Negation & \emph{If the user selects herself as a substitute, an error message is shown.} &
	\emph{A user \textbf{cannot select herself} as a substitute.} \\
	
	\bottomrule
\end{tabular}
\end{table*}


\begin{figure}[!t]
\center
\includegraphics[width=0.40\textwidth]{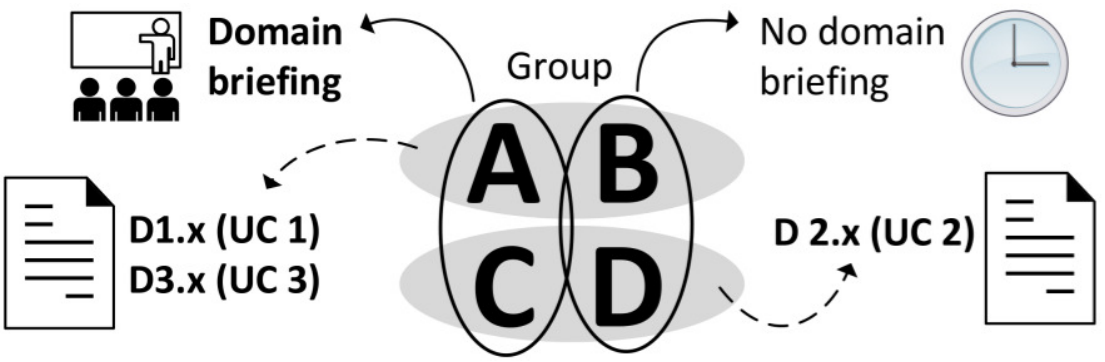}
\caption{Experiment Design Overview}
\label{fig:exp_groups}
\end{figure}



\begin{table}[!t]
\renewcommand{\arraystretch}{1.3}

\caption{Metrics for Evaluating Testing Impact}
\label{tbl:exp_metrics}
\centering
\begin{tabular}{@{}p{2cm}p{6cm}@{}} \toprule
  \textbf{Metric} & \textbf{Description} \\ 
	\midrule
	
	\emph{Total}$_{\texttt{Grp},\texttt{Def}}$ 				& Number of rated test cases regarding defect(s) \texttt{Def}, limited to participants from \texttt{Grp}\\
	\emph{Correct}$_{\texttt{Grp},\texttt{Def}}$			& Number of correct test cases regarding defect(s) \texttt{Def}, limited to participants from \texttt{Grp}\\
	\emph{Detected}$_{\texttt{Grp},\texttt{Def}}$ ~(D\,1.x only) 		& Number of test cases which detected \texttt{Def}, limited to participants from \texttt{Grp} \\
	\emph{Omitted}$_{\texttt{Grp},\texttt{Def}}$ ~(D\,2.x\,\&\,3.x only) 		& Number of test cases which omitted to test for the requirement specified by \texttt{Def}, limited to participants from \texttt{Grp} \\
	\emph{R\_Correct}$_{\texttt{Grp},\texttt{Def}}$	&   \emph{Correct}$_{\texttt{Grp},\texttt{Def}}$ $/$ \emph{Total}$_{\texttt{Grp},\texttt{Def}}$ \\
	\emph{R\_Detected}$_{\texttt{Grp},\texttt{Def}}$	&   \emph{Detected}$_{\texttt{Grp},\texttt{Def}}$ $/$ \emph{Total}$_{\texttt{Grp},\texttt{Def}}$\\
		\emph{R\_Omit}$_{\texttt{Grp},\texttt{Def}}$	&   \emph{Omitted}$_{\texttt{Grp},\texttt{Def}}$ $/$ \emph{Total}$_{\texttt{Grp},\texttt{Def}}$\\
	\emph{Independence}$_\texttt{Def}$ & Independence between the presence of defect \texttt{Def} in the use case and the correctness respectively omission in the inferred test cases, expressed as the p-value of Pearson's $\chi^2$ test (alt.\ hypotheses: \textbf{H}$_{A,C}$ resp.\ \textbf{H}$_{A,C})$  \\

	\bottomrule
\end{tabular}
\end{table}


\subsection{Experiment Design}
We inject pre-defined defects into real-world use cases and ask experiment participants to specify system test cases that appropriately verify the stakeholders' requirements. To this end, we provide tabular templates to be filled out within a 45 minutes time slot. No questions are allowed during the experiment. Additionally, we ask the participants about how difficult they perceived the inference of test cases for each use case, both quantitatively (8 point Likert-scale) and qualitatively using open questions.

To evaluate the research questions, we randomly assign participants to one of four Groups A-D (see Fig.\ref{fig:exp_groups} and Tbl.~\ref{tbl:exp_defects}):

\emph{For RQ\,2-1}, we inject two defects into UC1: An obviously incorrect defect D1.1, which requires to show an error message in case of success, and a more subtle defect D1.2, which suggests that a certain strictly manual check is executed automatically. Groups A \& B are faced with this flawed use case, whereas Groups C \& D serve as control group.

\emph{For RQ\,2-2}, we convert five positively stated requirements in UC2 and UC3 into their negative versions (D2.x and D3.x), carefully preserving the meaning of the each statement. Here, Groups A \& B receive the negated version of UC3 and serve as control group for UC2, and Groups C \& D receive the negated version of UC2 and serve as control group for UC3.

\emph{For RQ\,2-3}, we provide the participants of Groups A \& C with certain knowledge about the domain, directly before assigning the task to them: This briefing includes the purpose of the overall system, the relevant business processes, important rationales and necessary constraints from the perspective of a long-term employee of the company. In particular, the briefing includes the intended behavior for the defects D1.1 and D1.2. At the end of the presentation, questions are allowed to further foster the participants' understanding.
For this RQ, Groups B \& D serve as the control group without domain knowledge.

\subsection{Study Objects}
In order to keep the setting close to reality, we reuse a real-world SRS from an industrial partner. The original requirements specification was 21 pages long and written in natural language. It contained an overview, problem statement, supported business process description, functional requirements (organized as use cases) and non-functional requirements. 
For the experiment, we selected three out of 18 use cases, together with the original overview description and problem statement. All company-specific terms and acronyms were either removed or renamed due to legal reasons.

\subsection{Data Collection and Analysis Methodology}
We evaluate the obtained test cases for each defect by manual inspection: We assign \texttt{correct} to a test case if and only if the test explicitly covers the stakeholder's intended requirements (correct versions in Tab.~\ref{tbl:exp_defects}), \texttt{flawed} if unintended requirements are tested, and \texttt{omit} if the test does not cover the requirement at all. 
For semantic defects (RQ\,2-1), we furthermore inspected whether the participants actually \texttt{detect} the defects (i.e. are aware of it). We evaluate this based on whether we encountered remarks on D1.1 or D1.2 in the test cases, use cases or open questions. All metrics used for analysis are listed in Tab.~\ref{tbl:exp_metrics}, and we apply statistical tests for the following (alternative) hypotheses:

\begin{description}
	\item[\textbf{H}$_{A,C}$] The presence of a defect in the use case and the correctness of the test cases (regarding this defect) are not independent.
	\item[\textbf{H}$_{A,O}$] The presence of a defect in the use case and the omission of the correspnding requirement in the test cases are not independent.
	\item[\textbf{H}$_{A,D}$] The perceived difficulty is different for use cases with defects present. 
\end{description}

For the impact on the test quality, we test \textbf{H}$_{A,C}$ and \textbf{H}$_{A,O}$ (D\,2.x and 3.x only) using Pearson's $\chi^2$ test for each defect. To evaluate the impact on efficiency, we apply the Mann-Whitney test to \textbf{H}$_{A,D}$ in order to evaluate the impact on efficiency.  Both times, we demand a significance level of $\alpha$=0.05.
Furthermore, to validate the direction of the impact, i.e.\ whether it is indeed negatively for quality defects, we consult \emph{R\_Correct} and \emph{R\_Omit} for correct and flawed use cases.

For the relevance of domain knowledge (RQ\,2-3), we compare the aforementioned metrics, based on whether they received the domain knowledge briefing before the experiment.

\subsubsection*{Validity Procedures}
To ensure the reliability of the manual inspection, a second researcher independently rated 11 (25\%) randomly selected test cases. The obtained inter-rater reliability measures attested a very high level of agreement: we agreed on 93\% of all cases, and interpret Cohen's $\kappa=0.90$ as almost perfect agreement. Furthermore, to avoid unintentional influence of groups A and C beyond explaining the business process and employee experiences, the presentation was voice-recorded and checked later on, e.g., for accidental hints about the test cases.

\begin{table}[!t]
\renewcommand{\arraystretch}{1.3}

\caption{Experiment Results (sig.\ level $\alpha=.05$)}
\label{tbl:exp_results}
\centering
\begin{tabular}{@{}llrrlrrlrr@{}} \toprule
  \multicolumn{2}{c}{\textbf{Defect}}
	& \multicolumn{2}{c}{\textbf{Independence}}
	& 
	& \multicolumn{2}{c}{\textbf{R\_Correct}$_{\texttt{ID}}$} 																				
	&
	&  \multicolumn{2}{c}{\textbf{R\_Det.}/\textbf{R\_Omit}} \\ \cmidrule{3-4} \cmidrule{6-7} \cmidrule{9-10}
	
	ID & Grp
	& Omit & Corr.\ & 
	& Correct & Flawed & 
	& Correct & Flawed \\ \midrule
	

\multirow{3}{*}{D\,1.1} & All 
	& - & \cellmod 0.01 && 1.00 & 0.47 && - & 0.80 \\
& A\&C & - & 0.18 && 1.00 & 0.50 && - & 0.83 \\ 
& B\&D & - &  0.07 && 1.00 & 0.44 && - & 0.78 \\ 
\addlinespace

\multirow{3}{*}{D\,1.2} & All 
	& - & \cellmod 0.00 && 1.00 & 0.00 && - & 0.00 \\
& A\&C & - &\cellmod 0.01 && 1.00 & 0.00 && - & 0.00 \\ 
& B\&D & - &\cellmod 0.00 && 1.00 & 0.00 && - & 0.00 \\

 \midrule 

{D\,2.1} & All 
	& 0.54 & 0.88 && 0.83 & 1.00 && 0.68 & 0.53 \\ \addlinespace 

{D\,2.2} & All 
	& 0.20 & - && 1.00 & 1.00 && 0.26 & 0.53 \\ \addlinespace 

{D\,2.3} & All 
	& 0.28 & 0.71 && 1.00 & 0.80 && 0.47 & 0.71 \\

\midrule

{D\,3.1} & All 
	& 0.61 & 0.34 && 0.56 & 0.80 && 0.25 & 0.42 \\ \addlinespace 

{D\,3.2} & All 
	& 0.22 & - && 1.00 & 1.00 && 0.08 & 0.35 \\

\midrule

D\,2.x & All 
	& 0.15 & 0.91 && 0.90 & 0.93 && 0.38 & 0.51 \\
~~\& & A\&C & 0.19 & 0.83 && 0.84 & 0.90 && 0.34 & 0.51 \\ 
D\,3.x & B\&D & 0.59 & 1.00 && 0.96 & 0.95 && 0.42 & 0.50 \\


\bottomrule											
\end{tabular}
\end{table}

\subsection{Results and Interpretation}

We conducted the experiment as part of our RE lecture at the Technical University Munich (TUM). 
Participants were mostly advanced undergraduate or early graduate students of computer science or information systems, and the experiment was conducted late in the term so that students had a fundamental understanding of the contents and applications of SRS. 

\begin{figure}[!t]
\centering
\includegraphics[width=0.45\textwidth]{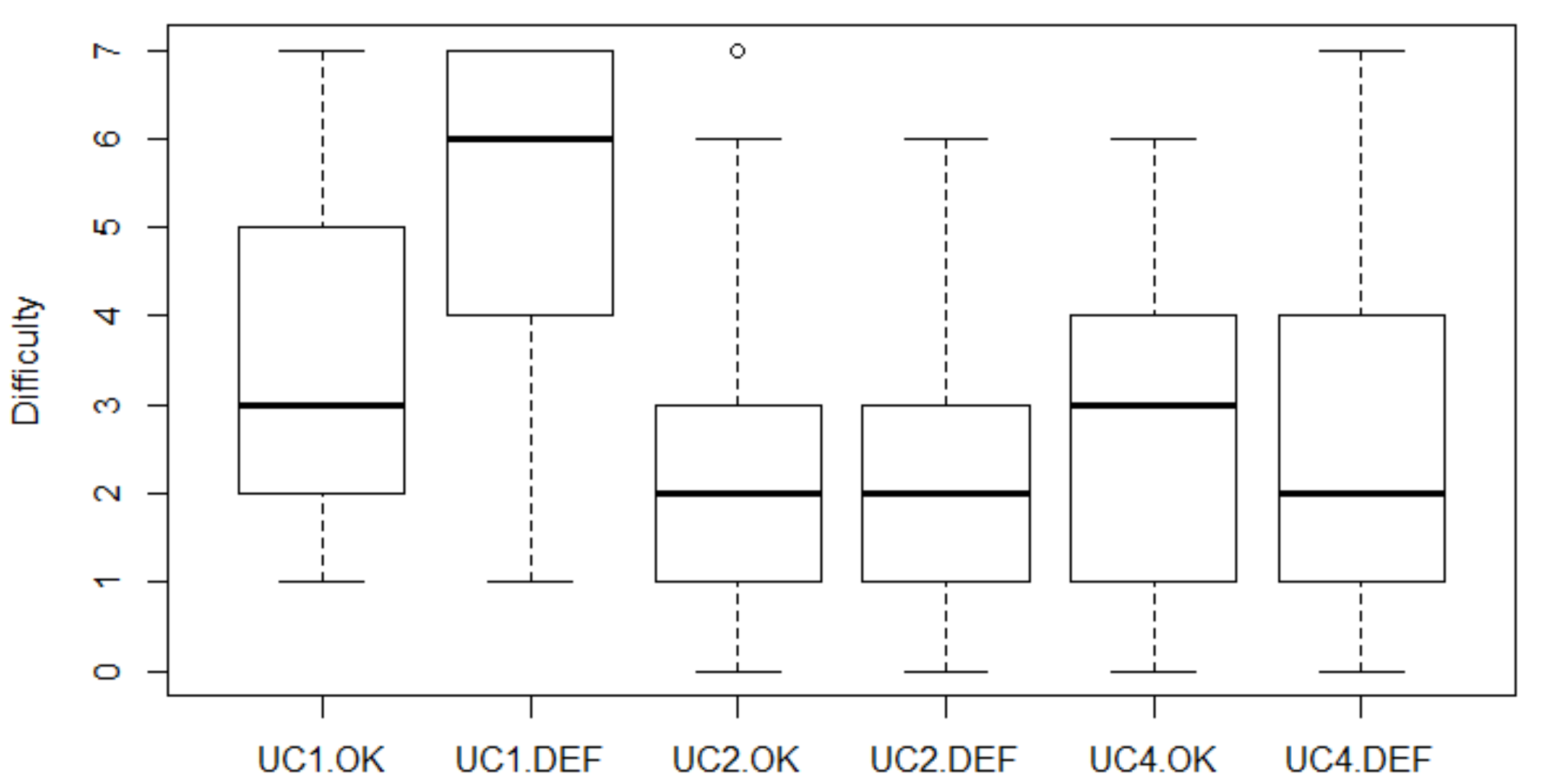} 
\caption{Difficulty per use case as perceived by participants}
\label{fig:exp_diff_conf}
\label{fig:exp_difficulty}
\end{figure}

Overall, 41 students participated in the experiment, with about ten students in every group. Tab.~\ref{tbl:exp_metrics} presents the obtained metrics and Fig.~\ref{fig:exp_diff_conf} illustrates the results of the self-evaluation of the participants. 

\subsubsection*{RQ\,2-1: Impact of Incorrect Statements}

The correctness of the test cases obtained from flawed specifications differed substantially between the defects D1.1 and D1.2: while for D\,1.2 no participant detected the defect and hence no correct test case was inferred, about half of all participants ($47\%$) explicitly corrected the defect D\,1.1 although 80\% of the participants actually recognized it. Hence, in 20\% of the cases, the defect was not detected at all. 
In contrast, for both D1.1 and D1.2, in the control group, a correct specification also led to correct test cases.
Therefore, we were able to reject the null hypothesis in favor of \textbf{H}$_{A,C}$ with statistical significance ($p \leq \alpha$ = 0.05 for both, D\,1.1 and D\,1.2). Furthermore, participants perceived the task as more difficult for UC\,1 with defects D\,1.1 and 1.2 present, hence rejecting the null hypothesis in favor of \textbf{H}$_{A,D}$ (p=0.02).
\paragraph*{Interpretation} The presence of incorrect statements in the SRS does indeed impact system testing. 
This is true regarding the quality of the obtained test cases, for obvious defects (D\,1.1) and even more for less obvious ones (D\,1.2). In addition, the (perceived) difficulty indicates an increase in required efforts, and hence also impacts efficiency of testing.

\subsubsection*{RQ\,2-2: Impact of Negated Statements}

In contrast to incorrect use cases (RQ\,2-1), the use of negative statements did not impact correctness of test cases substantially. For any negative statement but D\,2.3, correct test cases were obtained at least as often as for the non-negative statement. Consequently, we were unable to reject the null hypothesis in favor of \textbf{H}$_{A,C}$ for any negated statement. However, test cases did omit the specified requirement considerably more often ($+13\%$) but not significantly (\textbf{H}$_{A,O}$, p=0.15). Participants perceived the task equally difficult (\textbf{H}$_{A,E}$, p=0.46 (UC\,2) and p=0.95 (UC\,3)). Although not directly related to negative statements, participants expressed their dissatisfaction with the pragmatic quality of the specification as qualitative feedback, e.g., complaining about \emph{wording of requirements}, \emph{lack of chronological structure} and the \emph{use of passive sentences}.

\paragraph*{Interpretation} Our results suggest that negative statements do not impact testing in the sense that faults in terms of incorrect information are introduced into test cases, nor does it make the process of inferring these test cases more difficult. However, we observed a noticeably but statistically insignificant increase in omission of requirements expressed using negative statements, which might potentially lead to lesser quality in test cases due to untested requirements. 

\subsubsection*{RQ\,2-3: Relevance of Domain Knowledge}

For both incorrect as well as negative statements, results of participants introduced to the underlying business process (groups A\,\&\,C) did not vary considerably compared to the control group B\,\&\,D. Notably, participants were unable to detect or correct D\,1.2 (mandatory manual checks by a case handler).

\paragraph*{Interpretation} We were surprised that, although we explicitly mentioned the correct behavior in D\,1.2 in the briefing, participants were unable to compensate the defect. Within the experiment, the test engineers' knowledge of the system under consideration does neither compensate for incorrect requirements in specifications nor affects the quality of test cases inferred from specifications extensively using negative statements. Therefore, we conclude that, to a large extent, defects can propagate through the engineering process, even when people are briefed about the correct requirements.
 
\subsection{Threats to Validity}
We see three threats as limitations of this experiment: First, we relied on RE students as participants. Therefore, the obtained test cases were of rather poor quality in general, and certainly not at the level of experts in the field of system testing. 
Second, a briefing cannot lead to the same depth of domain knowledge compared to own experiences and observations over a prolonged amount of time, which we expect to be superior, e.g., in terms of recognition and trust. We thereby need to extend the experiment with industry experts in the future. Last, we only investigated selected defects and, thus, we have to be careful to generalize results to classes of SRS quality~\cite{lindland1994understanding}, especially concerning pragmatic quality factors.

\section{Discussion: Relevance of SRS Quality}
\label{sec:discussion}

In this section, we discuss our study results considering our introductory stated questions. 

\subsection{Under which circumstances does SRS quality matter?}

A generic answer to this question is provided by the two studies: the survey indicates to a (context-specific) extensive use of the SRS in only about half of all projects, and the experiment suggests that certain quality factors do and certain quality factors do not spread to subsequent results. Our survey provided first indicators to circumstances which correlate with (i)~a more extensive use of the SRS, e.g. the application domain of safety critical systems, whereas we could not confirm other circumstances proposed in literature. So far, our results indicate to two new challenges.

The first challenge is to determine the degree to which project circumstances may impact the \emph{needs} to document an SRS to a certain degree. If following the taxonomy of Gorschek and Davis~\cite{gorschek2008requirements}, we can see criteria of other dimensions beyond RE\footnote{For instance, criteria resulting from multi-project environments or from the socio-economic context of a customer including also cultural, psychological and even political facets.}, which all 
simply may not permit to identify stronger correlations in practice today. Moreover, some circumstances might not be equally important to others and there might be circumstances that dominate others. Our data set already indicates, for example, that safety/security-relevant systems can be considered as a \emph{dominant} circumstance in the sense that its presence dominates secondary circumstances such as the team-size. One potential explanation is that legal regulations enforce a rigor documentation no matter of secondary effects. However, although we could observe stronger correlations when considering non-safety-critical projects only, the number of projects was too small to draw meaningful conclusions.

The second challenge is to determine the degree to which project circumstances may impact the \emph{possibilities} to document an SRS to a certain degree. For instance, the experiment simulated the unavailability of stakeholders by not permitting questions during the assignment. Yet, we noticed that students, working on UC\,1 (incorrect statements, defects D\,1 and 2) in particular, tried to ask questions in the beginning nevertheless. Also, one participant explicitly stated 
\emph{[she] would like to be able to ask questions [to the customer]}. 

Both the needs to document an SRS to a certain degree and the possibilities that both arise from the characteristics of a project ecosystem need further investigations. 

\subsection{What dimensions of SRS quality matter?}

In general, one may argue that the dimensions of semantic and pragmatic quality of SRS, as proposed by Lindland et al.~\cite{lindland1994understanding}, become more important if requirements are documented to a larger extent, respectively used more extensively for communicating requirements. Therefore, the revealed correlations between criteria and SRS-based documentation/communication (cf.~Sec.~\ref{sec:survey}) are indicators for the relative importance for the semantic and the pragmatic quality as well. However, our experiment yields first insights into the absolute impact of quality: The defect D\,1.1, i.e. an incorrect statement is easily recognizable, leads to flawed test cases for about every second participant, and no correct test case could be derived from the less obvious defect D\,1.2. Although not part of the experiment, we do not expect better results for semantic defects in terms of missing information in the SRS. Therefore, we advocate a generalization for the semantic quality. That is, the semantic quality of the SRS is generally essential for subsequent engineering activities. 

However, the impact of the pragmatic quality appears to be more diffuse. While we could show indicators that negated statements do not impact engineering activities, we refrain from generalizing our results to pragmatic quality in general. We may even assume that negated statements are rather easily correctable compared to other pragmatic quality issues, e.g., the use of passive voice. In fact, qualitative feedback suggested that pragmatic quality was at least perceived as an obstacle, and our participants omitted requirements expressed in negated statements considerably more often. We believe that the pragmatic quality is rich on facets we do not yet properly understand with some factors having more severe impacts than others. Consequently, we strongly postulate to question and rigorosly investigate factors that best practice norms on (pragmatic) quality propagate. 

\subsection{How can we assure the quality of an SRS?}

Since quality assurance always comes at a certain cost, an immediate conclusion of our result is that SRS-based approaches applied independent of contextual circumstances are inherently inefficient. Based on the results and discussion, we advocate that quality assurance is not applied in general, but only were actually required. This can be achieved by different means, e.g., by introducing tailoring mechanisms or by using context-specific inductive approaches. However, independent of the actual mean, an efficient SRS-based quality assurance must include a decision procedure which specifies when the quality of an SRS needs \emph{not} to be assured. To this end, we hope the identified correlations of RQ\,2-1 and RQ\,2-2 will provide valuable first insights.

Finally an effective quality assurance must be able to assess the quality of the SRS in a way that is meaningful for the engineering endeavor. To this end, our results provide first indicators. On the one hand, the semantic quality of the SRS strongly impacts the outcome of subsequent activities, and consequently, SRS-based quality assurance can be very effective and must be considered during quality assurance. However, certain pragmatic quality factors that are proposed by best practices were not as influential as initially thought. Since semantic quality is difficult to assess, in particular for the predominant form of natural-language specifications, approaches providing reliable indicators for semantic quality based on syntactic properties (e.g., \cite{bernardez2004empirical}) seem promising.

\section{Conclusion}
\label{sec:conclusion}

We presented an investigation that showed how the relevance of SRS quality may depend on both project characteristics and what is considered as a quality defect. Therefore, efficient and effective quality assurance measures should consider applicability for specific contexts, not neglect semantic quality, and carefully select defects regarding SRS understandability.

\paragraph*{Relation to Existing Evidence}{
In \cite{lethbridge2003software}, Lethbridge et al. observed that documentation is frequently out-dated. In RQ\,1-1, we already discussed this might be one reason for the negative correlation between length of release cycles and SRS completeness. 
Interestingly, Lethbridge et al. also observed a moderate correlation ($0.43$, $p\leq$.05) between perceived accuracy and consultation frequency for requirements documentation, and, thus, one would suppose the SRS is used less for communication. However, the opposite appears in our result set ($\tau$=0.33, $p$=0.08). It is quite possible that in our study, we incidentally discovered a factor more important than perceived accuracy in long projects: the persistent nature of artifacts.
In a previous experiment, we could show that passive voice as another pragmatic quality factor \cite{femmer2014experiment} leads to difficulties in understanding sentences. This strengthens our confidence on the need to carefully evaluate quality factors, since some have and some don't have an impact on subsequent activities. }

\paragraph*{Future Work}{As future work, we will investigate both the needs to document an SRS to a certain degree and the possibilities that both arise from the characteristics of a project ecosystem. Furthermore, we believe that the pragmatic quality is rich on facets we do not yet properly understand with factors having more sever impacts than others, and we postulate the need to further investigate those factors propagated so far by best practice norms on (pragmatic) quality.}

\bibliographystyle{IEEEtran}
\bibliography{re-srs-relevancy}

\end{document}